\documentclass[prd,preprintnumbers,12pt]{revtex4}
\usepackage{epsfig}
\usepackage{graphicx}
\usepackage{bm}

\newcommand{\be}{\begin{equation}}

\newcommand{\ee}{\end{equation}}
\newcommand{\bea}{\begin{eqnarray}}
\newcommand{\eea}{\end{eqnarray}}

\newcommand{\nn}{\nonumber}
\newcommand{\de}{\partial}

\newcommand{\bi}{\begin{enumerate}}
\newcommand{\ei}{\end{enumerate}}
\newcommand{\bref}[1]{(\ref{#1})}
\newcommand{\A}{\alpha} \newcommand{\B}{\beta} \newcommand{\gam}{\gamma}
 \newcommand{\D}{\delta} 
\newcommand{\ep}{\epsilon}

        \newcommand{\s}{\sigma}
          \newcommand{\w}{\omega}
          
\newcommand{\h}{\eta}

\def\6{\partial}\def\7{\tilde}\def\8{\hat}

\def\pa{\partial}

\def\={{\;=\;}}\def\+{{\;+\;}}

\begin{document}
\preprint{ICCUB-11-49 \, UB-ECM-PF 11/42}

\title{Space-time transformations of the Born-Infeld gauge field of a D-brane}

\author{Roberto Casalbuoni}\email{casalbuoni@fi.infn.it}
\affiliation{Department of Physics, University of Florence and INFN Via G. Sansone 1, 50019 Sesto Fiorentino (FI), Italy}
\author{Joaquim Gomis}\email{gomis@ecm.ub.es}\affiliation{Departament d'Estructura i Constituents de la Mat\`eria}\affiliation{Departament de F\'isica,
Universitat de Barcelona, Diagonal 647, 08028 Barcelona, Spain and
C.E.R. for Astrophysics, Particle Physics and Cosmology, Barcelona
Spain\\and\\
Department of Theoretical Physics\\
Australian National University\\
Canberra, ACT 0200, Australia}
\author{Kiyoshi Kamimura}\email{kamimura@ph.sci.toho-u.ac.jp} \affiliation{Department of Physics, Toho University,
Funabashi, Chiba 274-8510, Japan}
\date{\today}

\begin{abstract}
Recently Gliozzi has shown that, under certain conditions, it is
possible to derive the Dirac-Born-Infeld action for an abelian gauge
field of a D-brane. Also, the action turns out to be invariant with
respect to a non-linear realization of the full Poincar\'e group. A
crucial role is played by the transformation properties of the gauge
field under the non-linear realization.
The aim of this note is to point out that these transformation properties  are derived directly from  the gauge fixing of the diffeomorphisms of the
 brane and the necessary compensating reparametrization when one performs a Lorentz transformation.
\end{abstract}


\maketitle


In a very recent paper \cite{Gliozzi:2011hj} Gliozzi has shown that it is possible to derive the Dirac-Born-Infeld (DBI) action for a gauge field lying on a  $d$-dimensional D-brane, by studying the non-linear representation of the Poincar\'e group in a
$D$-dimensional Minkowski space in terms of the $D-d$ transverse coordinates of the brane. A crucial point of the derivation lies in the transformation properties of the brane gauge field $A_a,~(a=0,1,\cdots,d-1)$. According to the author of reference \cite{Gliozzi:2011hj}: "Actually it appears to be no
 specific guiding principle (for deriving the transformations), nevertheless a trial and error method
 produced a surprisingly simple solution". Then, the author shows that these properties generate a non-linear realization of the Lorentz group on the gauge field. The aim of this brief note is to observe that the transformation properties of the gauge field follow simply from the requirement  of a gauge compensating transformation, necessary when one fixes the gauge identifying the longitudinal coordinates with the brane parameters. Notice that the gauge compensating transformation is also required in order to get the correct transformation properties for the transverse  coordinates (eq. (1) of
  \cite{Gliozzi:2011hj}).

We begin considering, as in  \cite{Gliozzi:2011hj},  a Poincar\'e
 invariant theory in a $D$-dimensional Minkowski space-time.  Furthermore we assume the presence of a stable $d$-dimensional extended object, a  D-brane.  We define  coordinates $X^\mu, (\mu=0,1,\cdots,D-1)$ in Minkowski space.
 $X^\mu(\sigma^\alpha)$ is  a target space vector, depending on the brane coordinates $\sigma^\alpha, (\alpha=0,1,\cdots,d-1)$. The covariant theory must be defined in such a way to be invariant under diffeomorphisms. Furthermore the coordinates
  transform under global Lorentz transformations as
 \be
 \delta_L X^\mu(\sigma^\alpha)={\omega^\mu}_\nu X^\nu,~~~~ \mu,\nu=0,1\cdots,D-1,~~~\alpha=0,1,\cdots,d-1.\ee
 We assume also the presence on the brane  of  a Born-Infeld (BI) $d$-dimensional abelian gauge field, $A_\alpha(\sigma^\beta)$ transforming as a scalar under the previous transformation
 \be\delta_L A_\alpha(\sigma^\beta)=0.\label{eq:2}\ee
 On the other hand, under local diffeomorphisms, $X^\mu$ and $A_\alpha$ should transform as
 a scalar and a vector respectively
 \bea
\D_D X^\mu(\s^\B)&=& {\ep^\B}\pa_\B X^\mu,\qquad 
\D_D A_\A(\s^\B)= {\ep^\B}\pa_\B A_\A+\pa_\A\ep^\B A_\B,\qquad \ep^\A=\ep^\A(\s^\B). \label{eq:3} \eea
Now we fix the gauge with respect to local diffeomorphisms by letting
\be
X^a(\sigma^\alpha)= \sigma^a \equiv x^a.\label{eq:4}\ee
Notice that the index $\alpha$ transforms under diffeomorphisms, whereas the index $a$ transforms under the Lorentz group. With this choice of gauge we identify the two kind of indices.

Then, in this gauge
\be X^\mu=(X^a(\sigma^b),X^i(\sigma^b))=(x^b,X^i(x^b)),~~~i=d,\cdots,D-1.\ee
The transverse coordinates $X^i(x^a)$ describe the long wave-length fluctuations of the brane, that is the Goldstone modes associated to the breaking of the translational invariance along the transverse direction specified by the index $i$.
Of course, also the Lorentz invariance $SO(1,D-1)$ is broken to $SO(1,d-1)\otimes SO(D-d)$, but this breaking does not produce any further Goldstone boson, see for example \cite{Gomis:2006xw}. This has to do with the inverse Higgs effect, see \cite{Ivanov:1975} and the more recent paper \cite{Low:2001bw}.
The Lorentz group $SO(1,D-1)$ is realized non-linearly in terms of the transverse coordinates $X^i(x^a)$.  We will show now how this comes about. First of all, let us notice that under transformations of $SO(1,d-1)$ the coordinates 
{ $X^a$} transform linearly
\be
\delta_L X^a={\omega^a}_bX^b.\label{eq:6}\ee
Then, let us consider Lorentz transformations  along the planes $(a,i)$, corresponding to the broken directions. In order to stay in the gauge chosen in \bref{eq:4} we have to perform at the same time a compensating diffeomorphism transformation such that
\be
0=\delta X^a|_{X^a=x^a}=(\delta_L X^a+\delta_D X^a)|_{X^a=x^a}= ({\omega^a}_i X^i+
\epsilon^b\de_bX^a)|_{X^a=x^a}={\omega^a}_iX^i(x^b)+\epsilon^a(x^b),\label{eq:7}\ee
where $\epsilon^a$ is the compensating diffeomorphism parameter,
\be
\epsilon^a(x^b)=-{\omega^a}_iX^i(x^b),\label{eq:7a}\ee
ensuring that the gauge is preserved. Now, the Lorentz transformation in the plane $(a,i)$ must be supplemented by the compensating diffeomorphism parameterized by $\epsilon^a$. For the transverse coordinates we get
\bea
\delta X^i(x^a)&=&{\omega^i}_a X^a+\epsilon^b\de_b X^i={\omega^i}_a~x^a-{\omega^b}_j~X^j~\de_bX^i\nn\\&
=&-{\omega^b}_j(\eta^{ji}x_b+X^j\de_bX^i).\label{eq:8}\eea
These transformations have been obtained, for example, in \cite{ ggrt,aksc,oamf} and discussed in \cite{Gliozzi:2011hj}.
The  identification of the indices $a$ and $\alpha$  implies that the gauge  field $A_a$  transforms linearly as a vector under $SO(1,d-1)$   and  undergoes an induced  Lorentz transformation in the plane $(a,i)$ due to the diffeomorphism necessary to keep the gauge fixed.
\bea
\D A_a(x)&=&{\ep^b}\pa_b A_a+\pa_a\ep^b A_b=
{-{\w^b}_i X^i(x)}\pa_b A_a-(\pa_a{\w^b}_i X^i(x)) A_b
\nn\\&=&
-{\w^b}_i \left(X^i~\pa_b A_a+(\pa_a X^i) A_b\right).\label{boostA}  \eea
This equation is the same as the equation (2) of \cite{Gliozzi:2011hj}. Therefore we have shown
that the transformation properties of $A_a$   have the same geometrical basis
 as the equation (\ref{eq:8}) expressing the transformation of the transverse coordinates,
 that is the need of the gauge compensating term. From this point of view the fact that the field
 $A_a$ supports a non-linear representation of the Lorentz group does not come as a surprise.

The transformations (6),  (7) in \cite{Gliozzi:2011hj}
of the world tensors $g_{\A\B}$ and  $F_{\A\B}$ are
also understood in the same manner since they transform under diffeomorphisms as
\bea
\D g_{\A\B}&=&\ep^\gam\pa_\gam g_{\A\B}+(\pa_{(\A} \ep^\gam) g_{\gam\B)} \nn\\
\D F_{\A\B}&=&\ep^\gam\pa_\gam F_{\A\B}+(\pa_{[\A} \ep^\gam) F_{\gam\B]}
\eea
and they are invariant under transformations in the plane $(a,i)$.
 Here the small round (square) parentheses stay for symmetrization (antisymmetrization).
Then the gauge fixed  $(a,i)$ transformations are
\bea
\D g_{ab}&=&(-{\omega^c}_i X^i)\pa_c g_{ab}-{\omega^c}_i (\pa_{(a}X^i) g_{cb)} =
-{\omega^c}_i\left( X^i \pa_c g_{ab}+(\pa_{(a}X^i) g_{cb)}\right) \nn\\
\D F_{ab}&=&(-{\omega^c}_i X^i)\pa_c F_{ab}-{\omega^c}_i (\pa_{[a}X^i) F_{cb]}=
-{\omega^c}_i\left( X^i \pa_c F_{ab}+(\pa_{[a}X^i) F_{cb]}\right).
\eea
We would like also to stress the fact that the transformation of $A_a$ could have been obtained thinking to it as it would be a "matter field" by following the standard techniques of the non-linear representations.

This can be seen in a more clear way by considering the general transformation of $X^\mu$ consisting in a translation plus a generic Lorentz transformation plus a diffeomorphism:
\be
\D X^a(x)={\omega^a}_b X^b+{\omega^a}_i X^i+ \rho^a+{\ep^\B}\pa_\B X^a
\label{delXaP}\ee
where  $\rho^a$ are the parameters of the translations.
To obtain the compensating diffeomorphism, we also require the variation of  $X^a$ to be zero in the gauge
$X^a(\s)=x^a, (a=0,...,d-1) $ . It gives
\be  \ep^a(x)=-({\omega^a}_i X^i+{\omega^a}_b x^b+ \rho^a).  \ee

The corresponding general transformations for $X^i, A_a$ are given by
\bea
\D X^i(x)&=&
-{\w^b}_j\left(
\h^{ji} x_b  {+ X^j}\pa_b X^i\right)+{\omega^i}_j X^j+\rho^i\,-({\omega^b}_c x^c
+\rho^b)\pa_bX^i,\label{delXiP}\eea
\bea
\D A_a(x)
&=&
-{\w^b}_i \left(X^i\pa_b A_a+(\pa_a X^i) A_b\right)+ {\omega_a}^bA_b-( {\omega^b}_c x^c+\rho^b)\pa_bA_a
.\label{delAP}\eea

Note that, as expected, the gauge field transforms under longitudinal transformations.
The last second term of $\D A_a$ in \bref{delAP} is the rotation of $A_a$ and
the last term indicates the $ISO(d)$ transformation of $x^b$ in  $A_a(x)$.

The infinitesimal transformation \bref{delAP} can  be written  as
\be
A'_a(x^b+{\w^b}_i X^i+{\omega^b}_c x^c+\rho^b)-A_a(x)=(
 {\omega_a}^b-{\omega^b}_i(\pa_a X^i))A_b. \label{delAP6}
\ee
This transformation shows that the BI field is  a covariant field of the non-linear realization
of the Poincar\'e ${ISO(1,D-1)}/\left({ISO(1,d-1)\otimes SO(1,D-d)}\right)$, if we consider that $A_a$ transforms under the vector representation of  $SO(1,d-1)$  subgroup of the unbroken group. Therefore $A_a$ should not be  considered as a Goldstone of some new space-time  symmetry but rather as a covariant field of the non-linear realization \cite{Coleman:1969sm,Callan:1969sn},  of the Poincar\'e group in $D$ dimensions.  
\acknowledgments

Joaquim Gomis aknowledges Peter Bouwknegt  for the warm hospitality at the Australian National University.
We acknowledge partial financial support from projects FP2010-20807-C02-01,
2009SGR502 and CPAN Consolider CSD 2007-00042.

\end{document}